\documentstyle[prb,aps,multicol]{revtex}
\begin{document}

\def \begincol{\begin{multicols}{2}} \def\endcol{\end{multicols} }
\def \ntext {\narrowtext}
\def \begincol{ } \def \endcol{ }
\def \ntext{} 

\def \et{{\it et al.~}}
\def \be{\begin{eqnarray}}
\def \ee{\end{eqnarray}}

\def \br{{\bf r}}
\def \bk{{\bf k}}
\def \bA{{\bf A}}
\def \bB{{\bf B}}

\def \tT{\tilde{T}}

\newcommand{\eqn}[1]{\mbox{Eq.\,($\ref{#1}$)}}
\newcommand{\fig}[1]{\mbox{Fig.\,($\ref{#1}$)}}

\title{ Phase-transitions in isotropic extreme type-II
  superconductors} \author{S-K. Chin, A. K. Nguyen and A. Sudb\o}
\address{Department of Physics, Norwegian University of Science and
  Technology, Trondheim, 7034 Norway.}

\date{\today}

\maketitle
\begin{abstract}
  Using large scale Monte Carlo simulations on a uniformly frustrated
  3DXY model, we report a {\em first order} vortex lattice melting
  transition in clean, {\em isotropic} extreme type-II ($\kappa
  \rightarrow \infty$) superconductors.  This work clarifies an
  important issue: the unpinned vortex liquid is always {\it
    incoherent} with no phase coherence in any directions for all
  anisotropies.  Previous claims of a disentangled vortex liquid for
  isotropic superconductors based on simulations, are due to finite
  size effects. We explicitly show that the effective vortex-line
  tension vanishes precisely at the superconducting phase transition
  in zero magnetic field. This loss of line tension is accompanied by
  an abrupt change in the connectivity of the vortex tangle across the
  superconductor. We also obtain results indicating that the
  connectivity of the vortex tangle changes in a similar way even in
  finite magnetic field, and suggest that this could also be
  associated with a genuine phase-transition.

\end{abstract}
\begincol

\begin{multicols}{2}

\section{Introduction}
Since the discovery of copper oxide based high temperature
superconductors (HTSC) \cite{Bednorz86}, which are of the extreme
type-II variety, there has been great interest in their
phenomenological phase diagrams.  Abrikosov's mean-field
description\cite{Abrikosov57}, which is valid for conventional low
temperature superconductors, is expected to be modified by the strong
thermal fluctuations in HTSC. Extensive research in both theory,
numerical simulations, and experiments over the years have resulted in
a general consensus on some of the fundamental issues. The current
understanding of HTSC in a uniform magnetic field $\bB$ is as follows.
In the absence of any pinning disorder, the low temperature Abrikosov
vortex lattice phase melts into a vortex liquid via a first order
transition at the temperature $T_m$. The Abrikosov vortex lattice
phase is characterized by a transverse triangular crystalline order
and a finite longitudinal phase coherence.

However, recently there has been some debate about the nature of the
vortex liquid which the Abrikosov vortex lattice melts into as
temperature increases.  Numerous simulations using the 3DXY model
\cite{Nguyen98loop,Nguyen98boson,Hu98,Nguyen98newop},
London\cite{Nordborg96} and Lowest Landau Level
approximations\cite{Hu97} have indicated that the vortex liquid is
{\it incoherent}, i.e. the phase coherence or superfluid density in
any direction, is zero. The crystalline order and phase coherence are
destroyed simultaneously at the melting transition. This scenario has
been supported by experiments on high quality YBCO
crystals\cite{Lopez96}. Other simulations using the 3DXY model
\cite{Li_Teitel,Chen97,Ryu98} have suggested that the longitudinal
phase coherence persists above the melting transition and only
vanishes at a higher ``entanglement'' temperature $T_E$. In this
scenario, the vortex liquid at $T_m<T<T_E$ would be disentangled with
relatively straight vortex lines. For $T>T_E$, it was claimed that
larger thermal fluctuations causes the vortex lines to be entangled
with a concomitant loss of global phase
coherence\cite{Li_Teitel,Chen97,Ryu98}.

Previously, a vortex liquid with nonzero longitudinal phase-coherence,
or superfluid density, was found in simulations on an isotropic
system\cite{Li_Teitel,Ryu98}.  More recently \cite{Ryu98}, it has been
proposed that for large $B$, $T_m$ and $T_E$ merge into a single
transition whereas for small $\bB$ they are well separated. The
authors found that for an isotropic system, the crossover between the
two regimes is at a filling fraction $1/18<f <1/6$, where the precise
definition of $f\propto B$ will be given shortly. However, these
simulations were performed on rather small systems. In the present
work, we have performed similar simulations on a much larger system,
and found an exclusively incoherent vortex liquid down to $f=1/60$.
This implies that the only thermodynamically stable vortex liquid
phase, is one which has zero longitudinal superfluid density, with
full translational and rotational symmetries and zero phase coherence
in all directions. We are therefore led to the conclusion that the
recently discussed \cite{Ryu98} crossover between the phase-coherent
vortex liquid and the phase-incoherent vortex liquid is a numerical
artifact.

The organization of this paper is as follows. In Section II we first
briefly review the model used in the simulations, and the quantities
we calculate. In Section III we discuss our results and their
implications. In Section IV we provide a summary and conclusion.

\section{Model}

We use a uniformly frustrated 3DXY model on a cubic lattice to
describe an isotropic, extreme type-II superconductors in a magnetic
field. The London model for superconductors can be readily derived
from the phenomenological Ginzburg-Landau (GL) model with the
approximation that the amplitude of the local complex order parameter
$\psi=|\psi|\exp[i\theta]$, is fixed. The Hamiltonian $H$ of the
London model consists of degrees of freedom in the phase $\theta(\br)$
and the gauge vector potential $\bA_{vp}(\br)$ associated with the
magnetic induction $\bB$ inside the system, i.e.
$H=H[\{\theta(\br),\bA_{vp}(\br)\}]$.

For an isotropic extreme type-II superconductor, the penetration depth
$\lambda$ is much larger than the coherence length $\xi$ such that the
GL parameter $\kappa=\lambda/\xi\rightarrow \infty$. This means that
the magnetic fields surrounding the vortex lines strongly overlap with
one another giving a spatially smooth $\bB$. This condition is ensured
in the regime where $\bB > \bB_{c1}$.  In other words, the
fluctuations of $\bA_{vp}(\br)$ on the length scale of $\xi$ are
negligible compared to the fluctuations of $\theta(\br)$. Therefore,
we can further simplify the problem by dropping the degrees of freedom
of $\bA_{vp}(\br)$ from the Hamiltonian $H$, and fix $B$ equal to the
external applied magnetic field. The resulting Hamiltonian
$H=H[\{\theta(\br)\}]$ is the 3DXY model. This $\kappa\rightarrow
\infty$ (or frozen gauge) approximation has been widely used as a
phenomenological model for superconductors. Note that, within this
model, the system has no {\it magnetic flux lines}, since there can be
no tubes of confined magnetic flux when $\lambda \to \infty$. The
system only exhibits {\it vortex lines}.

In order to perform numerical simulations on the resulting model, we
discretize the model on a 3D cubic lattice with grid spacing $\xi$.
The dimensionless Hamiltonian of this lattice model is given by
\cite{Hetzel92,Chen97,Nguyen98boson}
\be H[\{\theta(\br)\}]=- J_0 \sum_{\br,\alpha=x,y,z}
\cos\bigg(j_\alpha[\theta(\br)] \bigg)
\label{eqn_hamiltonian}
\ee
where $J_0=\Phi_0^2\xi/4\pi^2\mu_0\lambda^2$ is the isotropic coupling
energy, $\Phi_0=\hbar\pi/e$ is the flux quanta and $\mu_0$ is the
permeability of a vacuum in SI units. The dimensionless vector $\br$
labels the position of an arbitrary grid point. We define the
gauge-invariant phase difference $j_\alpha(\br)$ as \be j_\alpha(\br)
\equiv \theta(\br+{\bf e}_\alpha)-\theta(\br)
-\frac{2\pi\xi}{\Phi_0}\int^{\br+{\bf e}_\alpha}_{\br} d\br'\cdot
\bA_{vp}(\br')
\label{eqn_current}
\ee where ${\bf e}_\alpha$ is the unit vector along the $\alpha$-axis.
The convention is that $j_\alpha(\br)$ ``flows'' from the grid at
$\br$ to $\br+{\bf e}_\alpha$. In this model, it is natural to define
a dimensionless temperature $\tT=k_BT/J_0$.  The magnetic induction
$\bB$ is conveniently represented by the filling fraction \be
f=\frac{B\xi^2}{\Phi_0}.
\label{eqn_filling_fraction}
\ee

In this paper, we perform a Monte Carlo simulation on an isotropic
system with $V=N_x\times N_y \times N_z$ number of grid points. The
aim of our numerical simulations is to identify possible thermodynamic
phases (i.e. vortex lattice or liquid, etc.) and phase transitions
(first order or continuous) or crossovers associated with them.
Specifically, we are interested in calculating the internal energy,
specific heat, structure factor, and helicity moduli. These
thermodynamic quantities and their physical significance will be
discussed next. In addition, we introduce and define a quantity $O_L$
which we denote the vortex-path probability, and discuss some of its
physical implications.

A first order transition is indicated by a $\delta$-function anomaly
in the specific heat, equivalently a discontinuous jump in the
internal energy. On the other hand, the hallmark of a continuous
transition is a jump in the specific heat, modified by fluctuation
contributions to the anomaly, and a continuous internal energy. The
internal energy per site is obtained by averaging $H$ in
\eqn{eqn_hamiltonian} over the thermal equilibrium states, normalized
by the total number of grids, i.e. $E=<H>/V$. We define a
dimensionless specific heat per site $C$ using the standard
fluctuation theorem of a classical system with Gibbs
distribution\cite{Binney_book}: \be C=\frac{<H^2>-<H>^2}{V(k_B T)^2}
\label{eqn_specific_heat}
\ee

A convenient and widely used quantity to probe the {\it global} phase
coherence of the system is the helicity modulus, which is proportional
to the second derivative of the free energy associated with
\eqn{eqn_hamiltonian} with respect to an infinitesimal phase
twist\cite{Li_Teitel,Nguyen98loop}.  On a square lattice, the
dimensionless helicity modulus $\Upsilon_\alpha$ can be written as \be
\Upsilon_\alpha = \frac{1}{V}\bigg<\sum_\br\cos j_\alpha
-\frac{1}{\tT} \bigg|\sum_\br \sin j_\alpha\bigg|^2\bigg>.
\label{eqn_helicity_modulus}
\ee $\Upsilon_\alpha$ measures the ``stiffness'' of $\theta$ with
respect to twisting along the $\alpha$-direction.  If
$\Upsilon_\alpha>0$ then $\theta$ is stiff in the $\alpha$-direction,
or more precisely, there is {\em global} phase coherence or
superconducting response in that direction.  By tracking the
temperature at which $\Upsilon$ goes to zero, one can determine the
superconducting-normal metal transition temperature of the system. In
the mixed phase $\Upsilon_\parallel>0$ and $\Upsilon_\perp=0$, where
the subscripts $\parallel$ and $\perp$ denote the directions
longitudinal and transverse to $\bB$, respectively.

The structure factor probes the transverse crystalline order of the
vortex system. We adopt a conventional definition
\cite{Nguyen98boson}: \be S(\bk_\perp)=\frac{1}{(fV)^2}
\bigg<\bigg|\sum_{\br} v_\parallel(\br)\exp(i\bk_\perp\cdot\br)
\bigg|^2\bigg> \ee where $\bk_\perp$ is a two-dimensional reciprocal
vector and $v_\alpha(\br)$ is the local vorticity measured on the dual
square lattice grid \cite{Nguyen98loop,Nguyen98boson}, composed of the
centers of every direct unit cell.  A crystalline ordered phase is
characterized by $S(\bk_0)>0$ (or Bragg peaks) where $\bk_0$ are the
{\it discrete} set of reciprocal lattice vectors associated with the
crystal structure. On the other hand, $S(\bk)$ for a phase with full
rotational invariance exhibits ring patterns.

At a fixed temperature, the equilibrium configurations are generated
by making random changes to $\theta(\br)$ at each grid-point via the
the Metropolis algorithm. This is equivalent to randomly changing all
the six $j_\alpha$ attached to each grid-point. To ensure conservation
of vorticity in each unit cell, $j_\alpha$ has to be in the range
$[-\pi,\pi)$. Addition of $\pm 2\pi$ shall be administrated to bring
$j_\alpha$ back into range at every Monte Carlo step. {\it This
  procedure introduces vortex-loops into the system. Such loops are
  the elementary topological excitation of the model}. Periodic
boundary conditions (PBC) are imposed on $j_\alpha$ such that
$j_\alpha(\br+N_\mu e_{\mu})=j_\alpha(\br)$ for $\mu = (x,y,z)$. More
details about the Monte Carlo procedure may be found in
Refs.\onlinecite{Nguyen98loop,Nguyen98boson}.

We define $O_L$ as the probability of finding a directed vortex path
threading the entire system transverse to the induction $B$, {\em
  without using the PBC along the field direction} \cite{Jagla96}. It
is obtained by computing the number $N_V$ of times we find {\it at
  least one} such path threading the system in any direction $\perp B$
in $N_P$ different phase-configurations, normalized by $N_{P}$, i.e.
$O_L=N_V/N_P$. The fact that $O_L=0$ implies that there is no
connected path of vortex segments that threads the entire system in
the transverse direction, without using PBC along the field direction
several times. Now, let $N^{\alpha}_L$ ($\alpha \in [x,y,z]$) denote
the areal density of connected vortex paths threading the system in
any direction, including the direction parallel to the induction. It
is clear that in the Abrikosov vortex lattice phase $O_L=0$, and
$N^z_L = B/\Phi_0$, while $N^x_L = N^y_L=0$. Thus, $N_L^{\alpha}$ is a
conserved quantity at fixed induction $B$. On the other hand, $O_L=1$
implies that $N^{x,y}_L > 0$, and the {\it total} number of vortex
paths threading the system in any direction scales with system size,
but undergoes thermal fluctuations. Therefore, $N_L^{\alpha}$ is no
longer a conserved quantity.

We propose the following scenario to interpret the change in $O_L$ and
$N_L^{\alpha}$. Number conservation uniquely identifies a
$U(1)$-symmetry, and hence the low-temperature phase of the {\it
  vortex}-system (the dual of the phase-representation of the
superconductor) exhibits explicit $U(1)$-symmetry, since $O_L=0$. At
high temperatures $O_L=1$, $N^{\alpha}_L$ is not conserved, and the
$U(1)$-symmetry is broken. A $U(1)$-symmetric phase cannot be
analytically continued to a $U(1)$-nonsymmetric one. The change in
$O_L$ from $0$ to $1$ could therefore signal a phase-transition, in
this case involving breaking a global $U(1)$-symmetry, in finite as
well as zero magnetic field. However, to substantiate such a claim,
one needs to argue that $O_L$ is related to a local order parameter of
the system. To this end, we note that it is possible to transcribe the
{\it vortex-part} of the Ginzburg-Landay theory in the phase-only
approximation in such a way that the vortex-part of the theory is
specified in terms of a local, complex matter field $\phi({\bf x})$,
and that the theory then {\it explicitly} exhibits a $U(1)$-symmetry
\cite{Kleinert}. In the Lattice London Model, corresponding to the
Villain-approximation to the Ginzburg-Landau theory, this symmetry is
therefore only {\it implicit}, or ``hidden".  The probability of
finding a connected vortex-path starting at a point $\bf{x}$ and
ending at point ${\bf{y}}$, $G(\bf{x},\bf{y})$, is given in terms of
the two-point correlation function of the matter field
$\phi({\bf{x}})$,
$G({\bf{x}},{\bf{y}})=<\phi^*({\bf{x}})~\phi({\bf{y}})>$
\cite{Kleinert}. $O_L$ may be viewed as a special case of $G$, and in
the thermodynamic limit corresponds to $\lim_{|{\bf{x}}-{\bf{y}}| \to
  \infty} G({\bf{x}},{\bf{y}})$. If $G({\bf{x}},{\bf{y}})$ is non-zero
in this limit, this {\it suggests} the possibility of having
$<\phi({\bf{x}})> \neq 0$, and hence a broken $U(1)$-symmetry.
Although this does not constitute a proof that $O_L$ is connected to a
local order parameter whose expectation value is associated with a
broken $U(1)$-symmetry, it seems to be suggestive of such a
phase-transition existing even at finite magnetic field. Note that the
above local matter field $\phi$ appears to be the dual field of a
complex order parameter appearing in a somewhat different independent
approach to the same problem by Te{\v s}anovi{\'c}
\cite{Tesanovic98cm}.

\section{Results and Discussion}

In this section, we discuss the results of our simulations on an
isotropic system with size of $120^3$ grid points, which is the
largest to date on an isotropic system. The system is subdivided into
multiple sections, and the Monte Carlo procedure is implemented in
parallel across the sections using 3D ``black and white''
decomposition\cite{Nguyen98newop}. The filling fraction considered are
$1/f=20,40,60$, and $\infty$. The system is cooled from high
temperatures. For each temperature, a typical run consists of 120000
Monte Carlo sweeps across the whole lattice, $1/3$ of that were used
for equilibration. Near phase-transitions, up to 600000 sweeps were
used.

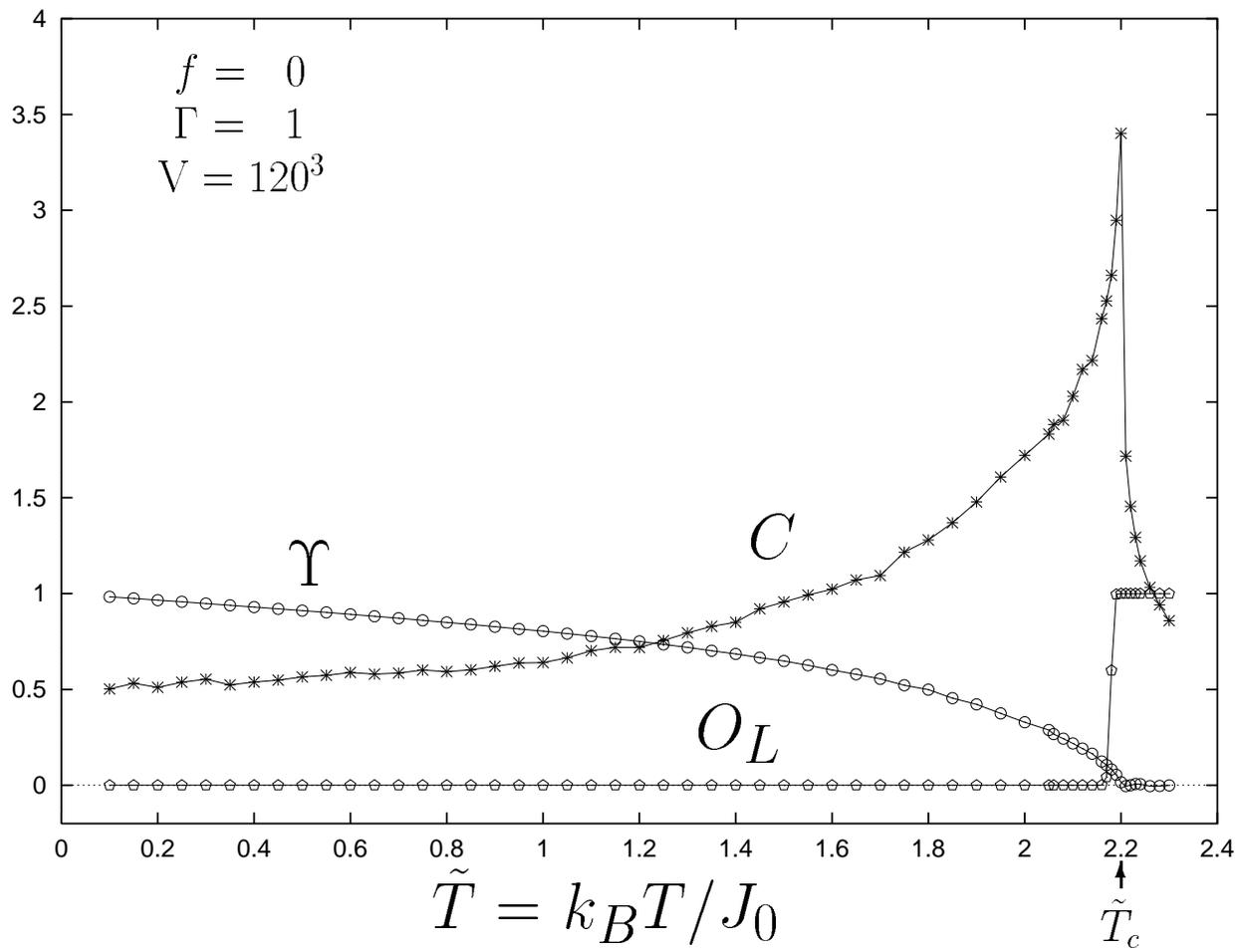
\begin{figure}
\caption{Specific heat $C$, superfluid stiffness $\Upsilon$, and vortex-path
  probability $O_L$ for $\Gamma=1$, $f=0$.}
\label{fig1} 
\end{figure}

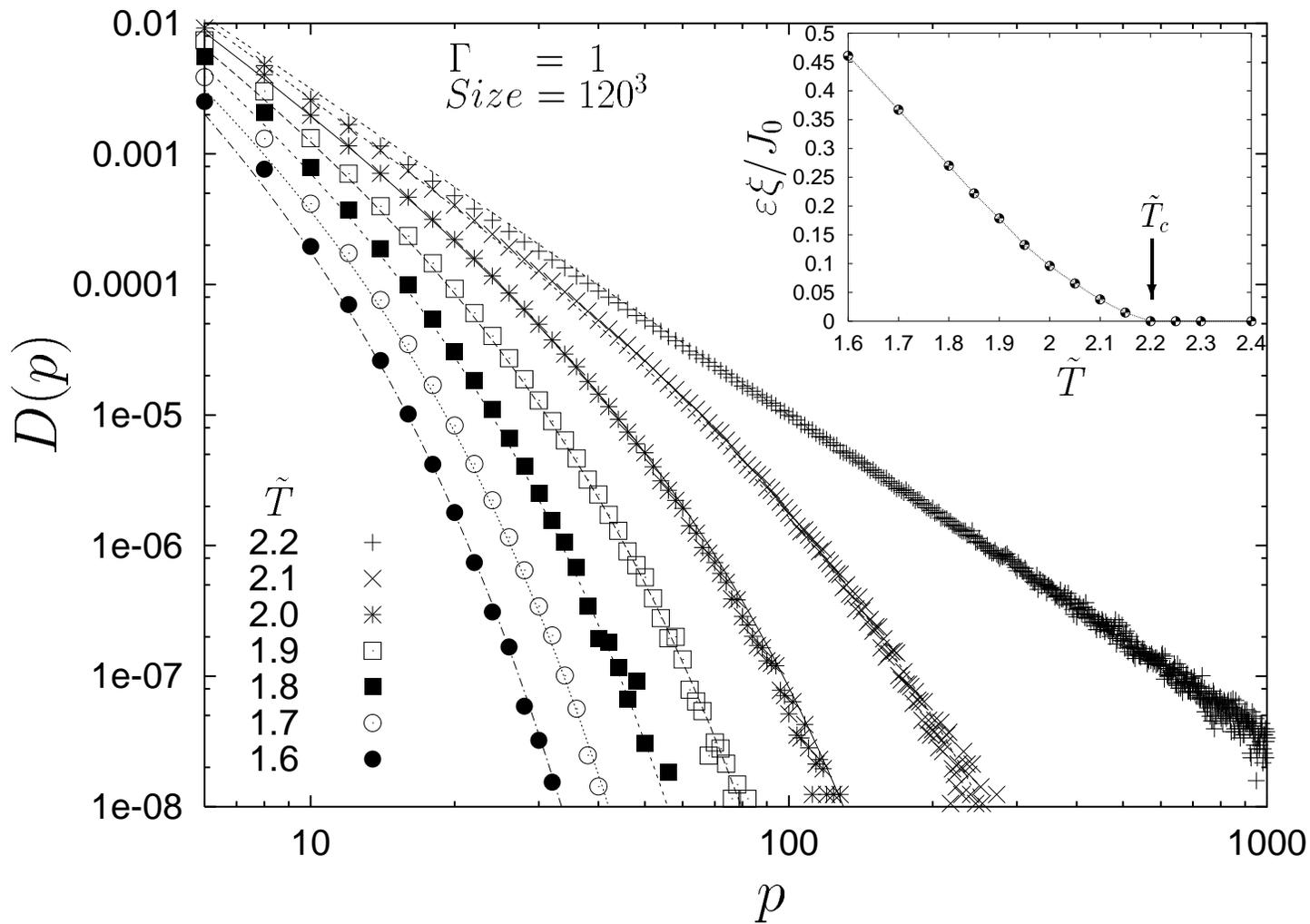
\begin{figure}
\caption{Vortex-loop probability 
  distribution $D(p)$ as a function of loop-perimeter $p$ for various
  temperatures, $\Gamma=1$, $f=0$. The lines in the figure are fits
  using $D(p)=A~p^{-\alpha}~\exp{(-\beta \varepsilon p)}$, with $A=1$,
  $\alpha=5/2$, and $\beta=1/k_B T$. $\varepsilon$ is the only fitting
  parameter in all plots. The inset shows the effective
  long-wavelength vortex-line tension $\varepsilon$. The solid line is
  a guide to the eye. $\varepsilon$ vanishes at $T=T_c$.  Here, $\xi$
  is the grid-spacing of the numerical lattice, and serves both as a
  unit of length and a measure of the superconducting coherence
  length.}
\label{fig2} 
\end{figure}

In \fig{fig1}, we present the results for the zero-field case, $f=0$,
where we show the specific heat $C$, the superfluid stiffness
$\Upsilon$, and the vortex-path probability $O_L$ as functions of
temperature. Note how the specific heat anomaly, the vanishing of the
superfluid stiffness and $O_L$, all coincide with $\tT_c \approx
2.20$. The physical interpretation of $O_L=1$ is that the effective
long-wavelength vortex-line tension vanishes
\cite{Nguyen98boson,Nguyen98newop,Tesanovic98cm}.  This claim is
substantiated by calculating the probability distribution $D(p)$ of
vortex loops as a function of perimeter $p$, at various temperatures.
We may fit this distribution to the form \cite{Hoye98pc}
\begin{eqnarray}
  D(p) & = & A ~ p^{-\alpha} ~ e^{-\beta \varepsilon p};~~~T < T_c
  \nonumber \\ & = & A ~ p^{-\alpha};~~~T \geq T_c
\end{eqnarray} 
where $\varepsilon$ is an effective, {\it temperature dependent},
long-wavelength line tension, $\beta = 1/k_B T$, and $\alpha = 5/2$.
The results are shown in \fig{fig2}, demonstrating that at $T=T_c$, a
purely algebraic decay is realized, implying $\varepsilon=0$.  Below
$T_c$, the exponential decay is well fitted, showing that $\varepsilon
\neq 0$.  The inset of \fig{fig2} shows $\varepsilon$ as a function of
temperature, as obtained in our simulations. A similar method of
extracting $\varepsilon$ for the zero-field case using a similar form
for $D(p)$ (without he power-law prefactor) has previously been used
by Li and Teitel, in Ref. \onlinecite{Li93}.

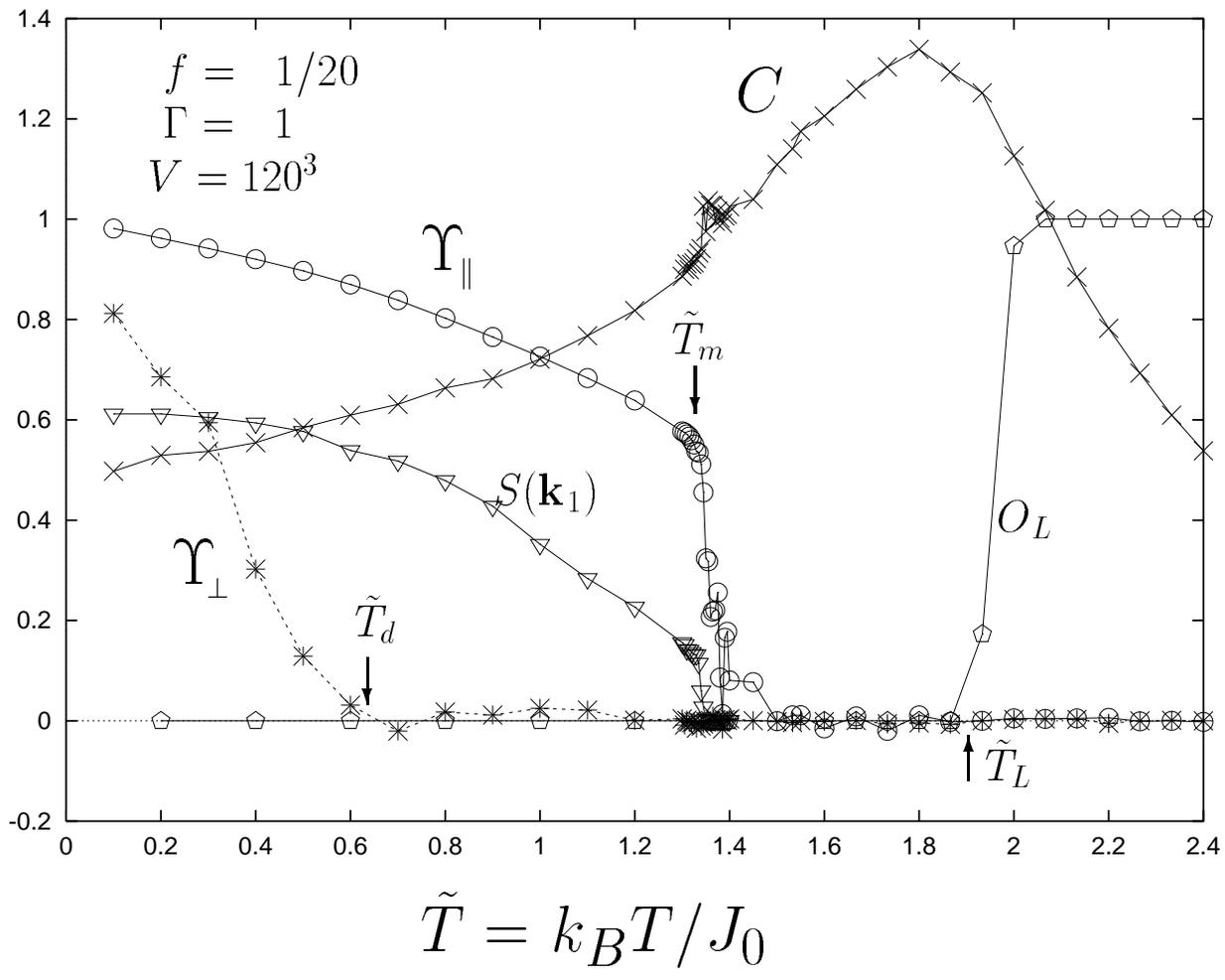
\begin{figure}
\caption{Specific heat per site $C$, helicity moduli $\Upsilon_\perp$ 
  and $\parallel$, vortex-path probability $O_L$, and structure factor
  $S(\bk_1)$ (where $\bk_1$ is the wavevector for one of the first
  order Bragg peak) as a function of $\tT$ for a system size $V=120^3$
  and $1/f=20$.  The melting temperature $\tT_m\approx 1.34$ is marked
  by the sharp drop of $S(\bk_1)$. The coincidence of a sharp peak in
  $C$ at $\tT_m$ confirms that the melting phase transition is first
  order. $\Upsilon_\parallel$ also vanishes at $\tT_m$ indicating that
  the triangular vortex crystal melts into an incoherent vortex
  liquid. At $\tT \approx 1.90$, $O_L$ jumps from $0$ to $1$,
  signalling a $U(1)$-symmetry breaking, or equivalently a
  $3DXY$-transition. }
\label{fig3} 
\end{figure}

The results for the quantities $S(\bk)$, $\Upsilon$ and $C$ for
$1/f=20$ are shown in \fig{fig3}.  Similar qualitative features are
also found for the cases $1/f=40,60$. First, $S(\bk_\perp)$ exhibits
six-fold Bragg peaks at low temperatures (not shown).  This indicates
that the low temperature phase is a triangular vortex lattice phase.
The destruction of the vortex lattice structure is marked by a melting
transition at $\tT_m \approx 1.34$ where $S(\bk_1)$ drops sharply to
zero, where $\bk_1$ is the wavevector of one of the six first order
Bragg peaks.  The sharpness of the drop in $S(\bk_1)$ strongly
suggests that the transition is first order. This is confirmed by the
appearance of a $\delta$-function like peak in $C$ at the same
temperature $\tT_m$. Coincidentally, $\Upsilon_\parallel$ which is
finite at low temperatures, also drops sharply to zero at $\tT_m$.
{\it The isotropic system exhibits longitudinal superconductivity
  below $\tT_m$, but not above}. Moreover, based on our above
discussion in Section II, we interpret the rise in the quantity $O_L$
from $0$ to $1$ as signalling that the effective long-wavelength
vortex-line tension vanishes. Thus, the vortex liquid phase is divided
into two regions in phase-space. In one region, the vortex liquid is
phase-incoherent i.e.  has no longitudinal superfluid density but has
finite vortex-line tension. We {\it propose} that this phase exhibits
a vortex-associated $U(1)$-symmetry. In the other region the vortex
liquid is phase-incoherent but has zero vortex-line tension.  We
propose that this phase exhibits a broken $U(1)$-symmetry.

Based on this, one can conclude that a first order melting transition
of the Abrikosov vortex lattice exists in an {\it isotropic} system in
the absence of pinning. The entire vortex liquid phase is {\it
  incoherent}, i.e. a vortex liquid phase with no longitudinal
superconductivity. Note that the same conclusion has been reached in
earlier simulations on {\it anisotropic}
systems\cite{Hu98,Nguyen98boson,Nguyen98newop}.  There have been
earlier reports of a disentangled vortex liquid, i.e. vortex liquid
with non-zero longitudinal phase coherence.  These results have been
obtained in similar simulation\cite{Li_Teitel,Chen97,Ryu98}.  In Ref.
\onlinecite{Ryu98}, it was argued that a phase-coherent vortex liquid
should be most easily observed in isotropic systems for $f\le 1/18$.
However, these simulations were performed on comparatively small
systems, typically of sizes no larger than $24^3$. Our present results
are based on much larger systems, $120^3$. We have observed a slight
difference in the temperatures at which $S(\bk)$ and
$\Upsilon_\parallel$ vanish in simulations on smaller system or lower
number of sweeps.  Therefore, it may be concluded that the existence
of a vortex liquid with a nonzero longitudinal superfluid density is a
numerical artifact of small system sizes and/or insufficient
simulation time \cite{Olsson98}.  It is well understood that the
$\lambda$-transition is driven by proliferation of thermally excited
loops of all sizes.  In recent work by some of us, it was proposed
that the same mechanism is driving the first order melting transition
at low $B$ and a newly discovered continuous transition involving the
breaking of U(1) symmetry at large $B$\cite{Nguyen98newop}.

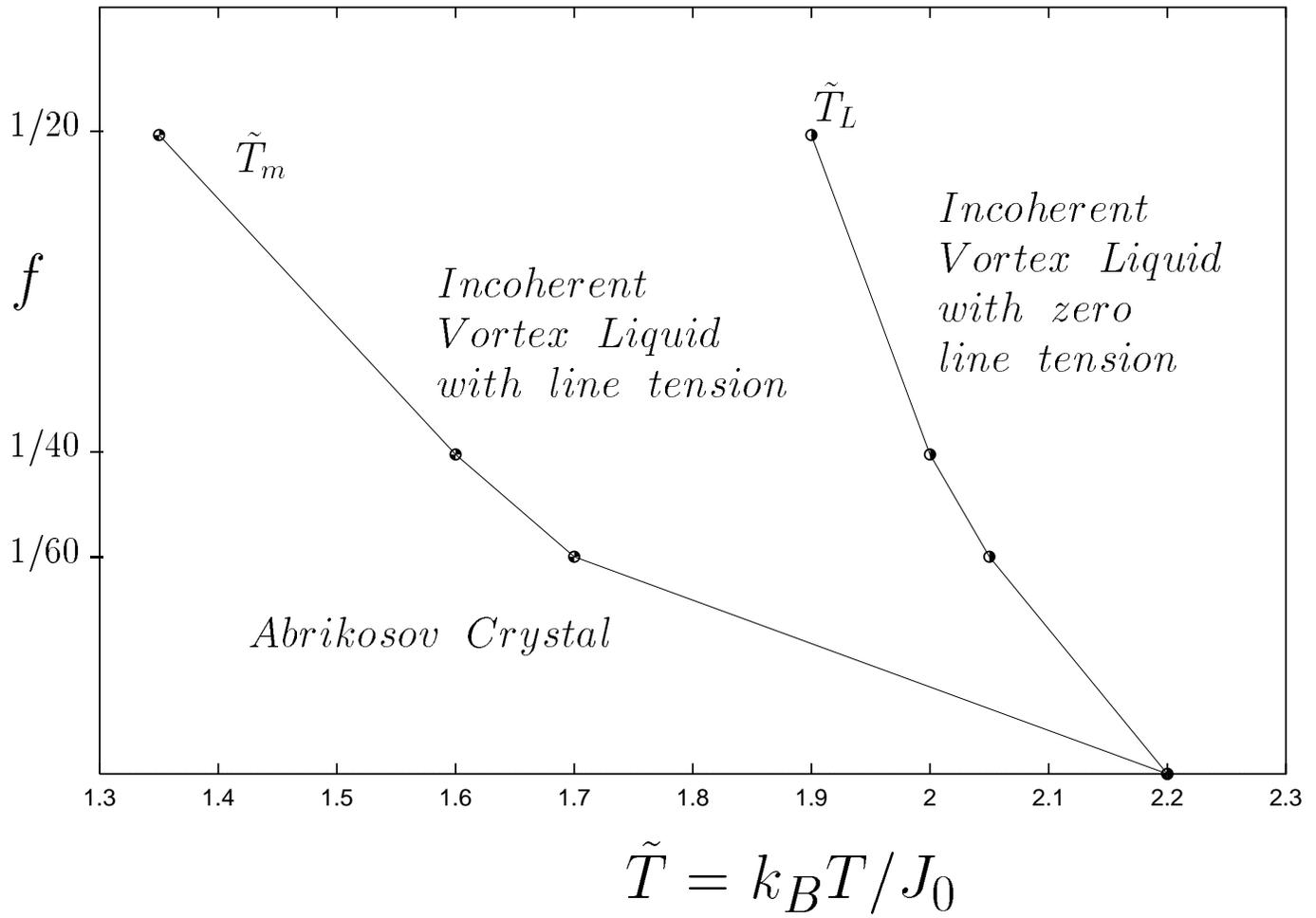
\begin{figure}
\caption{The intrinsic $f-\tT$ phase diagram of an isotropic system based on 
  simulations on system size $V=120^3$ and $f=0,1/20,1/40$ and $1/60$.
  The first order melting-line and the $3DXY$-line are denoted by
  $\tT_m$ and $\tT_{L}$, respectively.  }
\label{fig4} 
\end{figure}

Similar results for $1/f=40$ and $60$ enable us to propose a simple
phase diagram for an isotropic extreme type-II superconductor in the
absence of disorder (see \fig{fig4}). Immediately below and above the
$\tT_m$ line, the phases of the vortex system are identified as the
Abrikosov vortex lattice and the incoherent vortex liquid,
respectively.

In a pin-free system, one would expect $\Upsilon_\perp$ to be zero at
all temperature. In this case, the numerical lattice, on which
simulations are performed, effectively pins the vortex lines from
moving in the transverse plane and counteracts the Lorentz forces on
them in the presence of a transverse applied current. However, the
pinning is overcome by thermal fluctuations at higher temperature and
the vortex lines are depinned at temperature $\tT_d\approx 0.62$.
Fortunately, we see that $\tT_d\ll \tT_m$, which means that near
$\tT_m$, the vortex lines, and the melting process, are completely
free from the effects of the numerical grid. Therefore, the features
of $C$, $\Upsilon_\parallel$ and $S(\bk)$ at $\tT_m$ are genuine
thermodynamic effects.

\section{Conclusion}

We have performed simulations of the uniformly frustrated 3DXY model
on a large {\it isotropic} system ($120^3$ grid points) for a variety
of filling fractions $1/f=20,40,60$ and $\infty$. We found a first
order melting transition in this isotropic system for all the three
non-zero values of $f$ considered. The longitudinal phase coherence
and triangular crystalline order of the Abrikosov vortex crystal are
simultaneously destroyed at the melting transition. Above the melting
temperature, the incoherent vortex liquid is the only thermodynamic
phase. We have demonstrated that previous claims of the existence of
disentangled vortex liquid\cite{Li_Teitel,Chen97,Ryu98} is due to
performing simulations using insufficient system sizes and simulation
times.

We have shown that the effective vortex-line tension vanishes
precisely at the zero-field superconducting transition. The loss of
superfluid stiffness, the loss of line tension, and the abrupt change
in the connectivity of the vortex tangle, as signalled by the change
in the quantitity $O_L$ across the system, all coincide in this case.
A similar change in connectivity across the vortex system takes place
at finite magnetic field. The results of the Ref.
\onlinecite{Jagla96}, the present paper, and in particular those of
Ref. \onlinecite{Nguyen98newop} strongly indicate that this change in
connectivity is sharp in the limit of large systems, thus indicating
the loss of number-conservation of connected vortex paths threading
the system.  Since the finite probability of finding a connected
vortex path threading the system in a direction other than the
magnetic field may be tied to the finite expectation value of a local
complex matter field \cite{Kleinert}, this lends further support to
the argument that the change in $O_L$ signals the breaking of a
$U(1)$-symmetry \cite{foot1}. At the very least this proposition
appears to be intriguing enough to warrant further investigation.

We finally caution the reader that we so far have not been able to
detect any anomaly in specific heat at the suggested new finite-field
transition inside the vortex liquid, for the isotropic case. Even the
anomaly at the first order melting transition is weak in the isotropic
case, and is difficult to bring out in simulations.  It may be that
considerably larger systems are needed for the isotropic case in order
to see signals in the specific heat of the suggested new transition
due to the small amount of entropy in the transition. This is the
reason why the anomaly at the first order vortex lattice melting
transition is difficult to observe in simulations. It is conceivable
that increasing the anisotropy of the system should bring out the
anomaly clearer, if it exists. This indeed is the case for the anomaly
at the vortex lattice melting transition. A weak anomaly associated
with the putative $U(1)$-transition, may in fact have been observed
for the anisotropic case in Ref. \onlinecite{Nguyen98newop}.

\section*{Acknowledgment}
We thank J. S. H{\o}ye, P. Olsson, S. Teitel, and Z. Te{\'s}anovi{\v
  c} for discussions and consrtructive remarks. This work is supported
by the Norges Forskningsr\aa d under grants 115577/410, 110566/410 and
110569/410. We acknowledge the use of the Cray T3E-600 at the
Norwegian Supercomputing Project of NTNU, Trondheim, Norway.


\begin{thebibliography}{10}
\bibitem{Olsson98} Similar results as the ones we have obtained for
  the longitudinal phase coherence in the vortex liquid phase in this
  paper, have now been reported by P. Olsson and S. Teitel,
  cond-mat/9810169.
\end{thebibliography}

\end{multicols}
\newpage

\endcol

\end{document}